\documentclass{moriond}

\def\invfb{\mathrm{fb}^{-1}}
\def\kl{\kappa_\lambda}
\def\kTwoV{\kappa_{2V}}
\def\ttbar{\mathrm{t}\bar{\mathrm{t}}}

\newcommand{\Photo}{\includegraphics[height=35mm]{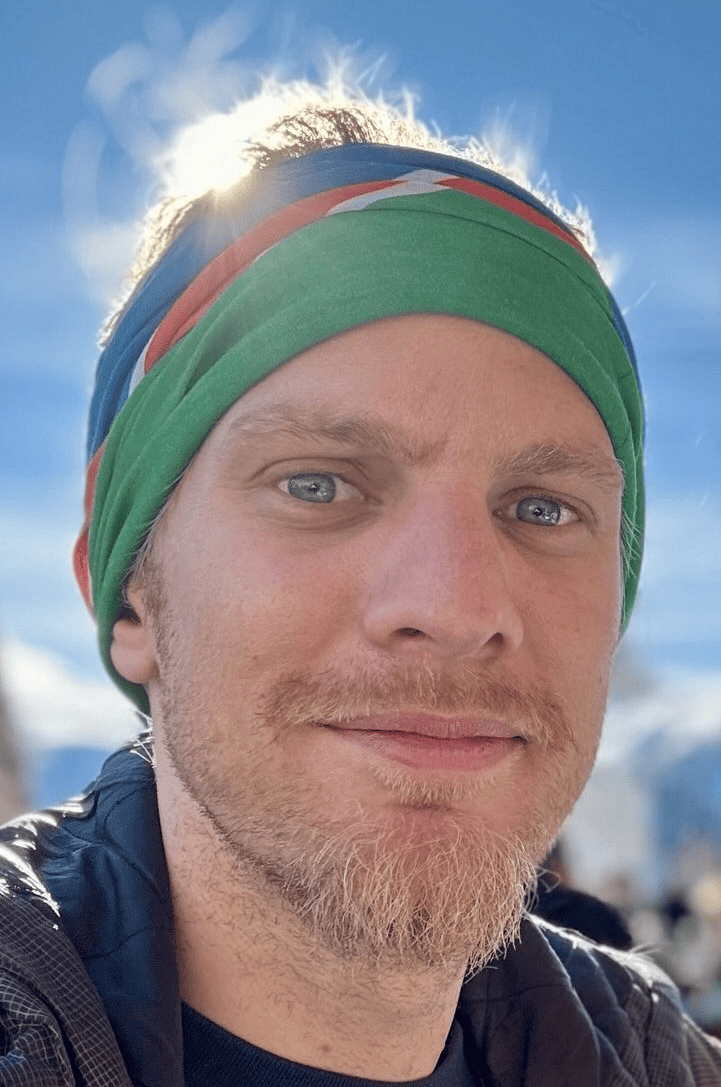}}

\begin{document}
\vspace*{4cm}
\title{Searches for Resonant and Nonresonant Multi-Higgs Production in ATLAS and CMS}

\author{Emmanouil Vourliotis\\
        on behalf of the ATLAS and CMS Collaborations~\footnote{Copyright 2026 CERN for the benefit of the ATLAS and CMS Collaborations. Reproduction of this article or parts of it is allowed as specified in the CC-BY-4.0 license}}

\address{University of California San Diego, CA, U.S.}

\maketitle\abstracts{
The measurement of the Higgs boson self-coupling is one of the key physics goals of the LHC program, providing direct access to the shape of the Higgs potential and offering sensitivity to physics beyond the Standard Model (BSM).
Higgs pair production probes the trilinear self-coupling, while triple Higgs production is sensitive to the quartic coupling.
In addition, resonant searches for heavy scalars decaying to pairs of Higgs bosons or to a Higgs boson and a new scalar are motivated by a broad class of BSM models.
Recent results on multi-Higgs production searches from the ATLAS and CMS experiments at the Large Hadron Collider (LHC) are presented, covering both resonant and nonresonant production modes across a variety of final states.
}

\section{Introduction}\label{sec:Intro}


The Higgs boson self-couplings are among the most important Higgs boson properties, as they determine the Higgs potential and are sensitive to beyond the Standard Model (BSM) effects.
They are also among the most challenging ones to measure experimentally.
The trilinear coupling can be probed in processes involving the simultaneous production of two Higgs bosons (``di-Higgs" production), with a cross section $10^3$ times smaller than that of the single Higgs production.
The quartic coupling is accessible through processes involving the simultaneous production of three Higgs bosons (``tri-Higgs" production), with a cross section that is $5 \times 10^2$ times smaller than the di-Higgs one.
In pursuit of these challenging measurements, the ATLAS and CMS Collaborations have been performing searches for multi-Higgs production.
Using the experience gained by analyzing the data from LHC Run 2 ($\sqrt{s} = 13~\mathrm{TeV}$), both collaborations have been publishing optimized multi-Higgs analyses, exploiting the recently recorded LHC Run 3 data ($\sqrt{s} = 13.6~\mathrm{TeV}$) and employing cutting-edge machine learning techniques to enhance their sensitivity.
Any deviations from the SM prediction could be explained by BSM heavy scalar resonances decaying to a Higgs boson pair or a Higgs boson and a new scalar resonance, and the ATLAS and CMS Collaborations are actively searching for such processes.

\section{Analysis Highlights}\label{sec:Analyses}


Both the ATLAS and CMS Collaborations have performed searches for di-Higgs production to the $\mathrm{bb}\gamma\gamma$ final state.
The ATLAS Collaboration performed such a search~\cite{ATLAS_HHbbgg} using a combination of the full Run 2 (2015--2018) and a partial Run 3 (2022--2024) data set, corresponding to $308~\invfb$ in total, making it the first ATLAS analysis to use 2024 data.
Boosted-decision-trees (BDTs) were trained in different kinematic regions defined based on the $\mathrm{bb}\gamma\gamma$ invariant mass, and their outputs were used to define signal regions (SRs).
The $m_{\gamma\gamma}$ distributions across 14 exclusive SRs were then fitted simultaneously.
The signal strength is measured to be compatible with the SM expectation, $\mu_\mathrm{HH} = 0.9^{+1.3}_{-1.0}\mathrm{(stat.)^{+0.6}_{-0.5}\mathrm{(syst.)}}$, with an observed(expected) significance of $0.84(1.01)\sigma$.
Upper limits on the signal strength are extracted: the observed limit is $\mu_\mathrm{HH} < 3.8$ while the expected limit assuming (no) HH production is $\mu_\mathrm{HH} < 3.7(2.6)$.
This analysis also sets constraints on the observed(expected) $\kl = 2.7(1.0)^{+2.3(+4.7)}_{-3.0(-1.7)}$ and $\kTwoV = 1.6(1.0)^{+0.6(+1.1)}_{-1.9(-1.3)}$ coupling modifiers, obtained by a two-dimensional fit where they are included as free parameters.

The CMS Collaboration made public an analysis~\cite{CMS_HHbbgg} in the same final state, using an early Run 3 data set (2022--2023) corresponding to $62~\invfb$.
Two complementary approaches were explored to extract the results: performing a fit only with the $m_{\gamma\gamma}$ distribution (1D-fit approach) and fitting simultaneously the $m_{\gamma\gamma}$ and $m_\mathrm{bb}$ distributions (2D-fit approach).
In both approaches, three ``resolved" SRs, requiring at least two small-radius jets, were constructed based on the output of multiclass multivariate analysis methods.
A ``boosted" SR, exploiting events with at least one large-radius jet, was also implemented to further improve the sensitivity.
Observed(expected) upper limits on the signal strength were obtained under the assumption of no HH production, yielding $\mu_\mathrm{HH} < 11(7.3)$.
Constraints were also placed on the $\kl$ coupling modifier, excluding values outside of the range $-6.1 < \kl < 12.8$.
The results above correspond to the 2D-fit approach and are found to be consistent with those of the 1D-fit approach within uncertainties.

\begin{figure}[!hbtp]
\centering
\includegraphics[width=0.44\linewidth]{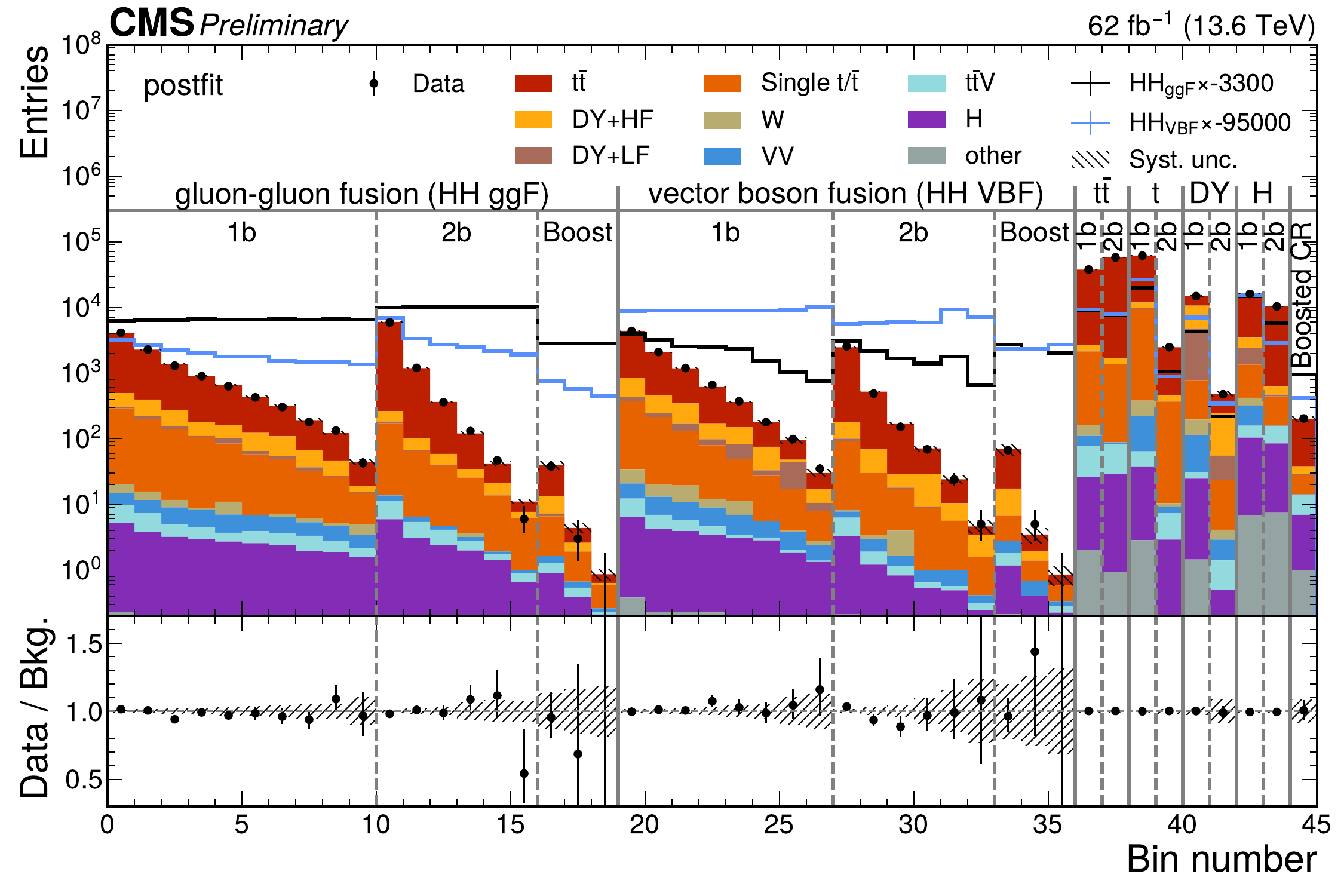}
\qquad
\includegraphics[width=0.30\linewidth, trim={0 30px 0 0}]{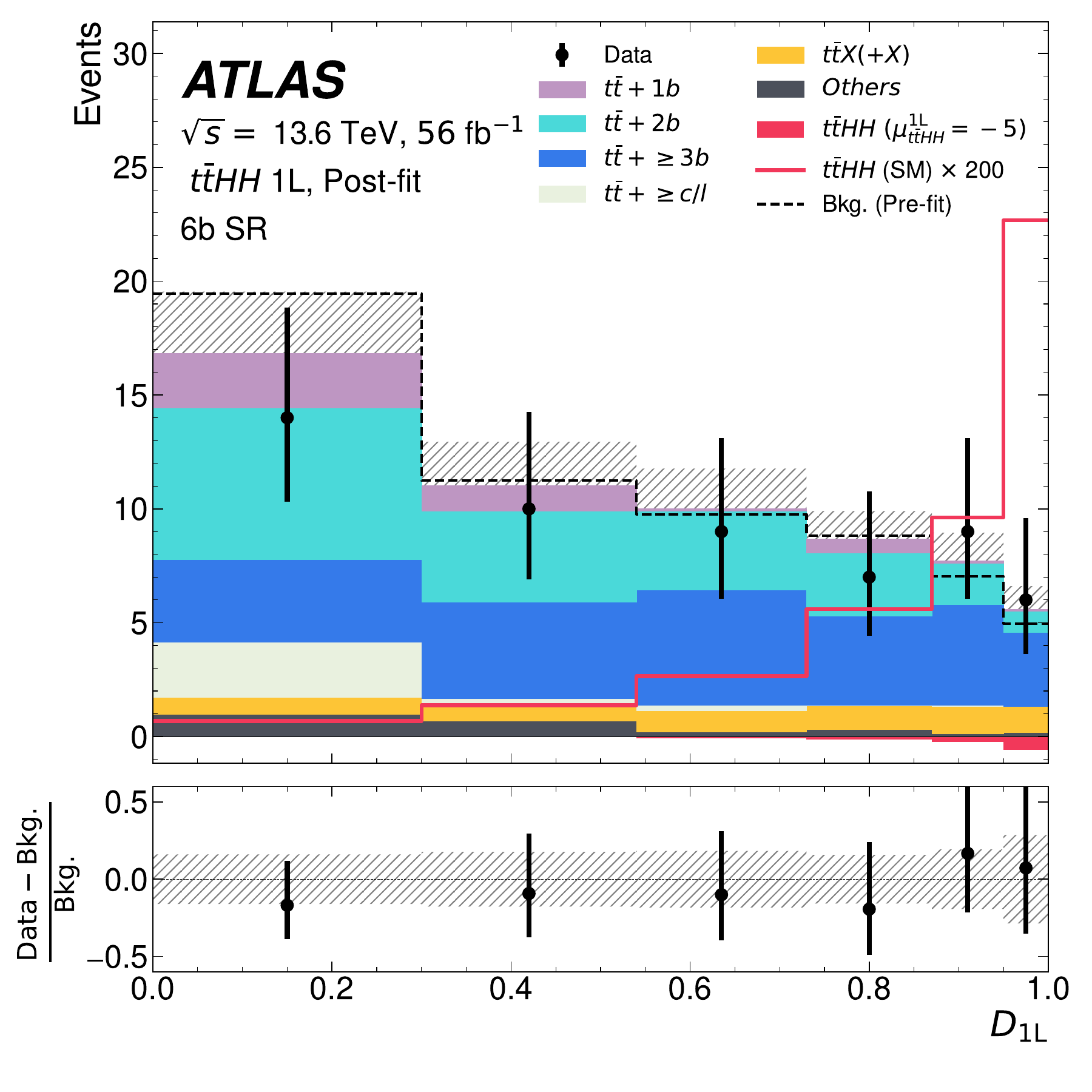}
\caption[]{Post-fit SRs and CRs of the CMS $\mathrm{HH} \rightarrow \mathrm{WWbb}$ analysis~\cite{CMS_HHWWbb} (left), and the 1L NN score of the ATLAS ttHH analysis~\cite{ATLAS_ttHH} (right).}
\label{fig:HHWWbb_ttHH}
\end{figure}

The di-Higgs production in the final state with two bottom quarks and two leptonically decaying W bosons was studied~\cite{CMS_HHWWbb} by the CMS Collaboration using $62~\invfb$ from the 2022--2023 data set.
This analysis relied on a series of binary and multiclass neural networks (NNs) to separate the signal and background processes.
To maximize sensitivity, the gluon-gluon fusion (ggF) and vector boson fusion (VBF) signals were further categorized in resolved and boosted SRs and based on the number of b-tagged jets in the event.
Figure~\ref{fig:HHWWbb_ttHH} (left) shows the ggF and VBF HH SRs together with the dedicated control regions (CRs) for the background processes ($\ttbar$, t, DY H) after their simultaneous fit to the data.
The analysis results in an observed(expected) upper limit on the signal strength of 12.7(18.6), and the $\kl$($\kTwoV$) coupling modifiers were constrained within the range $[-9.7, 15.8]$($[-0.27, 2.32]$).

The ATLAS Collaboration performed the first search for ttHH production~\cite{ATLAS_ttHH} using the 2015--2023 data set of $196~\invfb$.
Three complementary signatures were probed, targeting different combinations of the $\ttbar$ and HH decay modes:
a single-lepton (``1L") category targeting semileptonic $\ttbar$ with $\mathrm{HH} \rightarrow 4\mathrm{b}$, a same-sign or multilepton (``SSML") category targeting leptonic $\ttbar$ and $\mathrm{HH} \rightarrow \mathrm{b}\ell\ell$, and a ``$\mathrm{bb}\gamma\gamma$" category targeting $\mathrm{HH} \rightarrow \mathrm{bb}\gamma\gamma$.
For the final selection, transformer NNs were used for the 1L and SSML categories, while BDTs were employed for the bb$\gamma\gamma$ category.
The signal extraction has been performed by simultaneously fitting the outputs of the transformer models, shown for the 1L category in Figure~\ref{fig:HHWWbb_ttHH} (right), and the $m_{\gamma\gamma}$ distribution for the bb$\gamma\gamma$ category.
The analysis set an observed(expected) upper limit on the signal strength $\mu_\mathrm{HH} < 20(21)$ and constrained the Higgs effective field theory Wilson coefficient $c_{\ttbar\mathrm{HH}}$ in the range $[-3.9, 3.4]$.

\begin{figure}[!hbtp]
\centering
\includegraphics[width=0.33\linewidth]{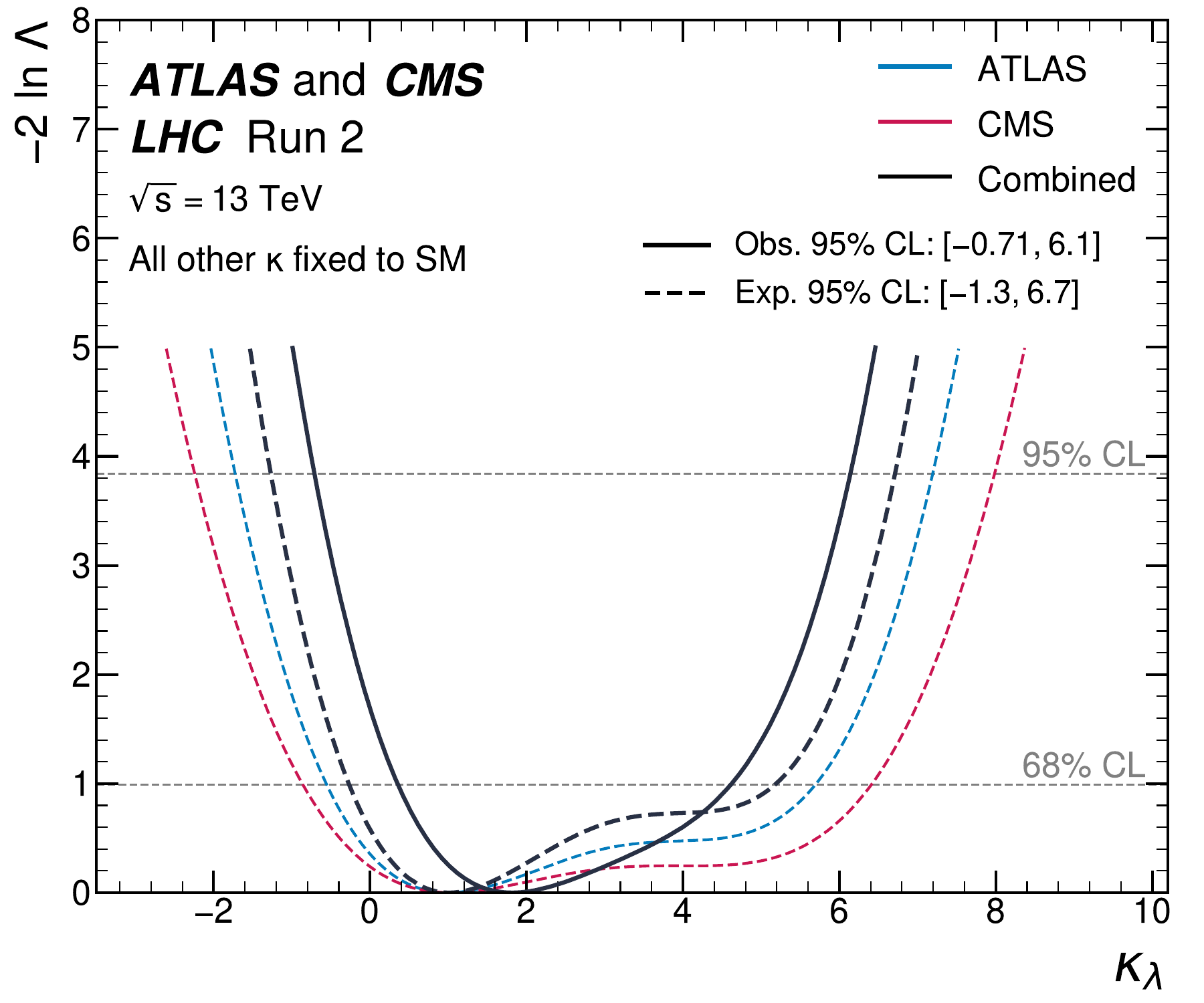}
\includegraphics[width=0.33\linewidth]{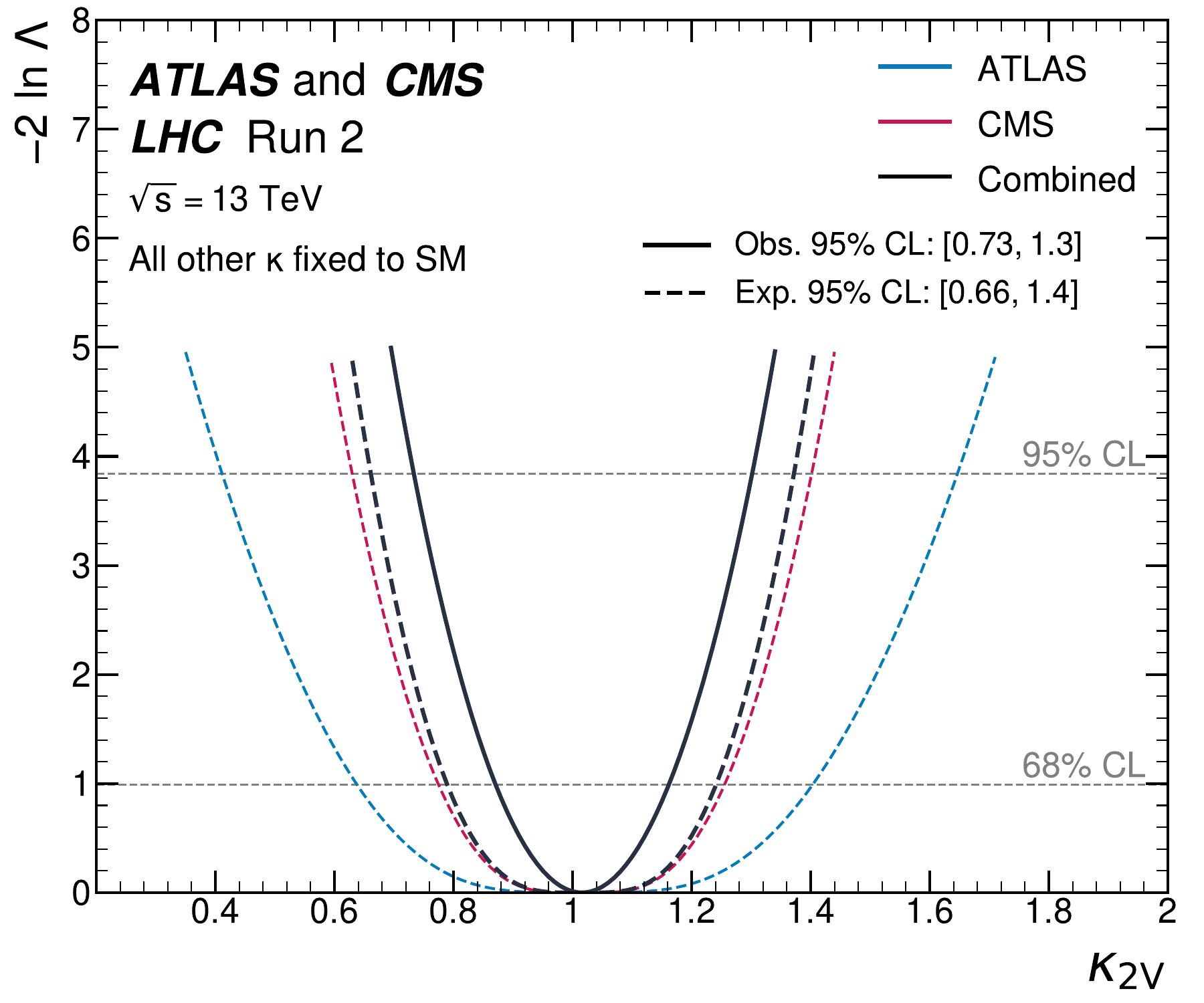}
\includegraphics[width=0.27\linewidth, trim={0 35px 0 0}]{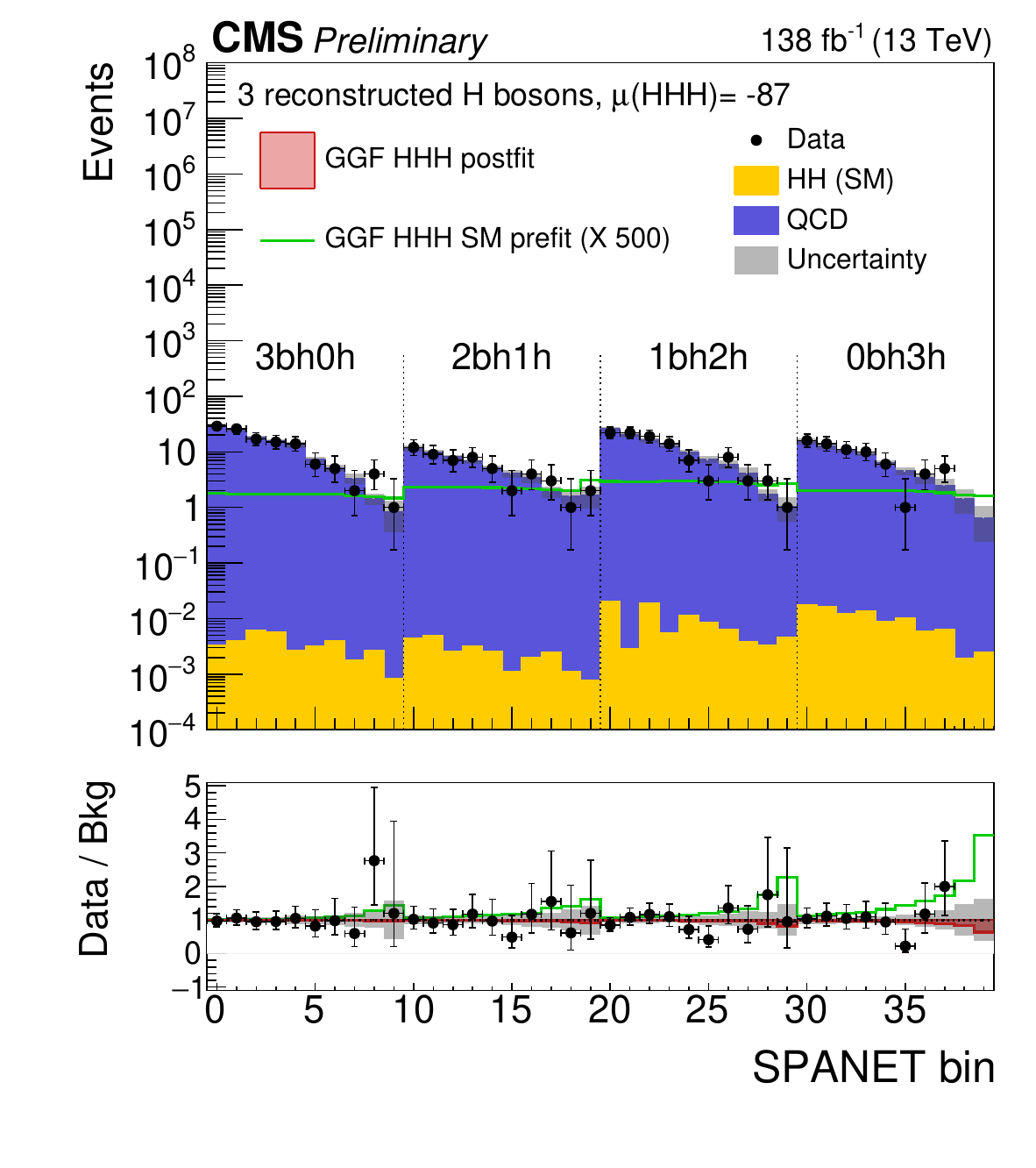}
\caption[]{$\kl$ (left) and $\kTwoV$  (middle) likelihood scans of the ATLAS and CMS HH combination~\cite{ATLASCMS_HHComb}, and the post-fit SRs of the CMS $\mathrm{HHH} \rightarrow 6\mathrm{b}$ analysis~\cite{CMS_HHH6b} (right)}
\label{fig:HHComb_triHiggs}
\end{figure}

The above analyses showcase only a small part of the effort ongoing for the search for the di-Higgs production, which is now focusing on Run 3 data.
The Run 2 analyses on the same topic have reached a level of maturity and completeness that led the ATLAS~\cite{ATLAS_HHComb} and CMS~\cite{CMS_HHComb} Collaborations to combine the analyses of their HH program, publishing a comprehensive set of results on di-Higgs properties.
Going a step further, the statistically independent analyses by both ATLAS and CMS were combined~\cite{ATLASCMS_HHComb} to push the sensitivity even more.
This combination resulted in the most stringent limits on HH production to date, $\mu^\mathrm{observed(expected)}_\mathrm{HH} < 2.5(1.7)$, and the tightest constraints on the $\kl$ and $\kTwoV$ coupling modifiers, as shown in Figure~\ref{fig:HHComb_triHiggs} (left).

Undertaking an even more challenging effort, the CMS Collaboration performed a search~\cite{CMS_HHH6b} for tri-Higgs production in the final state with six bottom quarks, using data corresponding to $138~\invfb$ of luminosity recorded during Run 2 (2016--2018).
Apart from the small tri-Higgs cross section, a further complication is that only $\sim\!25\%$ of the events have three reconstructed Higgs bosons ($\sim\!50\%$ and $\sim\!25\%$ with 2 and $\leq1$ Higgs bosons, respectively).
To overcome this difficulty, analysis events are classified as signal (HHH) and different background processes (HH, $\ttbar$, QCD) using SPANet~\cite{SPANet}.
An additional SPANet layer was employed to categorize the signal events into SRs according to the number of resolved (``h") or boosted (``bh") Higgs boson they contain.
The SRs, after their simultaneous fit to the data, are shown in Figure~\ref{fig:HHComb_triHiggs} (right).
This analysis gave the best limit to date on tri-Higgs production, $\mu^\mathrm{observed(expected)}_\mathrm{HHH} < 588(572)$, and the most stringent $\kappa_4$ coupling modifier constraint to date, $|\kappa_4| < 190$.

In addition to the nonresonant di-Higgs production, the ATLAS and CMS Collaborations have been searching for resonant production of a Higgs boson pair or a Higgs boson and a new scalar particle (denoted as ``S" in ATLAS and as ``Y" in CMS).
Most recently, the ATLAS Collaboration analyzed data from Run 2 (2015--2018, $140~\invfb$) and Run 3 (2022--2023, $59~\invfb$) looking for a heavy scalar resonance, ``X", decaying to the $\mathrm{X} \rightarrow \mathrm{S}(\mathrm{bb}) \mathrm{H}(\gamma\gamma)$ signature~\cite{ATLAS_XSHbbgg}.
To efficiently cover the full parameter space, which spans a large combination of mass hypotheses for the X and S particles, two parametrized NNs (PNNs) were used:
one in a region with one b-tagged jet (for the cases where one of the signal b-jets is not reconstructed) and one in a region with two b-tagged jets.
For the same purpose, additional mass hypotheses were probed using interpolation.
No significant excesses were reported, with the highest local significance being of the order of $2\sigma$.

Finally, the CMS Collaboration performed a search~\cite{CMS_XYHZZbb} for $\mathrm{X} \rightarrow \mathrm{H}(\mathrm{ZZ}) \mathrm{H}(\mathrm{bb})$ and $\mathrm{X} \rightarrow \mathrm{Y}(\mathrm{ZZ}) \mathrm{H}(\mathrm{bb})$ with the Run 2 data set (2016--2018), corresponding to $138~\invfb$ of luminosity.
In this analysis, parametrized BDTs were used in SRs with 0, 1 and 2 large-radius jets to probe the different $m_\mathrm{X}$ and $m_\mathrm{Y}$ hypotheses.
Upper limits were set on the production cross sections of the above processes, and those were found to be comparable to those of searches for the same process in typically more sensitive final states, such as $\mathrm{H} \rightarrow \gamma\gamma$ and $\mathrm{H} \rightarrow \tau\tau$.


\section{Outlook}\label{sec:Outro}

The ATLAS and CMS Collaborations build on the experience gained from Run 2 analyses and push the sensitivity for probing resonant and nonresonant multi-Higgs production.
Apart from the inclusion of the recent Run 3 data, significant progress has been made by employing improved strategies and advanced machine learning techniques.
This has led to updated projections~\cite{HLLHC} for High-Luminosity LHC, which report that the expected significance for the observation of di-Higgs production will exceed $7\sigma$ by the end of the Phase 2 running program.

\section*{References}
\bibliography{EmmanouilVourliotis}

@misc{ATLAS_HHbbgg,
      title         = "{Study of Higgs boson pair production in the $HH \rightarrow b \overline{b} \gamma \gamma$ final state with 308 fb$^{-1}$ of data collected at $\sqrt{s} = 13$ TeV and 13.6 TeV by the ATLAS experiment}", 
      author        = {ATLAS Collaboration},
      year          = {2025},
      eprint        = {2507.03495},
      archivePrefix = {arXiv},
      primaryClass  = {hep-ex},
      url           = {https://arxiv.org/abs/2507.03495},
      note          = {Submitted to Phys. Lett. B},
}

@misc{CMS_HHbbgg,
      title         = "{Search for HH production decaying into two b quarks and two photons in pp collisions at 13.6 TeV with a partial CMS Run 3 dataset}", 
      author        = {CMS Collaboration},
      year          = {2025},
      url           = {https://cds.cern.ch/record/2945062},
      note          = {\href{https://cds.cern.ch/record/2945062}{CMS-PAS-HIG-25-007}},
}

@misc{CMS_HHWWbb,
      title         = "{Search for Higgs boson pair production in the $\mathrm{b\bar{b}WW}$ decay channel with two leptons in the final state using proton-proton collision data at $\sqrt{s}$ = 13.6 TeV}", 
      author        = {CMS Collaboration},
      year          = {2026},
      eprint        = {2604.02127},
      archivePrefix = {arXiv},
      primaryClass  = {hep-ex},
      url           = {https://arxiv.org/abs/2604.02127},
      note          = {Submitted to JHEP}
}

@misc{ATLAS_ttHH,
      title         = "{Search for Higgs boson pair production in association with top-quark pairs using 196 fb$^{-1}$ of proton-proton collision data at $\sqrt{s}=$ 13 and 13.6 TeV with the ATLAS detector}", 
      author        = {ATLAS Collaboration},
      year          = {2026},
      eprint        = {2603.13113},
      archivePrefix = {arXiv},
      primaryClass  = {hep-ex},
      url           = {https://arxiv.org/abs/2603.13113},
      note          = {Submitted to JHEP}
}

@article{ATLAS_HHComb,
  title         = "{Combination of Searches for Higgs Boson Pair Production in $pp$ Collisions at $\sqrt{s}=13~\mathrm{TeV}$ with the ATLAS Detector}",
  author        = {ATLAS Collaboration},
  collaboration = {ATLAS Collaboration},
  journal       = {Phys. Rev. Lett.},
  volume        = {133},
  pages         = {101801},
  year          = {2024},
  doi           = {10.1103/PhysRevLett.133.101801},
  eprint        = {2406.09971},
  archivePrefix = {arXiv},
  primaryClass  = {hep-ex},
}

@misc{CMS_HHComb,
      title         = "{Combination of searches for nonresonant Higgs boson pair production in proton-proton collisions at $\sqrt{s}$= 13 TeV}", 
      author        = {CMS Collaboration},
      year          = {2025},
      eprint        = {2510.07527},
      archivePrefix = {arXiv},
      primaryClass  = {hep-ex},
      url           = {https://arxiv.org/abs/2510.07527},
      note          = {Submitted to Rep. Prog. Phys.}
}

@misc{ATLASCMS_HHComb,
      title         = "{Combination of ATLAS and CMS searches for Higgs boson pair production at $\sqrt{s} = 13$ TeV}", 
      author        = {ATLAS and CMS Collaborations},
      year          = {2026},
      eprint        = {2602.23991},
      archivePrefix = {arXiv},
      primaryClass  = {hep-ex},
      url           = {https://arxiv.org/abs/2602.23991},
      note          = {Submitted to PRL}
}

@misc{CMS_HHH6b,
      title         = "{Search for nonresonant triple Higgs boson production in the six b-quark final state in proton-proton collisions at 13 TeV}", 
      author        = {CMS Collaboration},
      year          = {2025},
      url           = {https://cds.cern.ch/record/2945361},
      note          = {\href{https://cds.cern.ch/record/2945361}{CMS-PAS-HIG-24-012}},
}

@Article{SPANet,
	title         = "{SPANet: Generalized permutationless set assignment for particle physics using symmetry preserving attention}",
	author        = {Alexander Shmakov and Michael James Fenton and Ta-Wei Ho and Shih-Chieh Hsu and Daniel Whiteson and Pierre Baldi},
	journal       = {SciPost Phys.},
	volume        = {12},
	pages         = {178},
	year          = {2022},
	doi           = {10.21468/SciPostPhys.12.5.178},
    eprint        = {2106.03898},
    archivePrefix = {arXiv},
    primaryClass  = {hep-ex},
}

@misc{ATLAS_XSHbbgg,
      title         = "{Search for a resonance decaying into a scalar particle and a Higgs boson in the final state with two bottom quarks and two photons with 199 fb$^{-1}$ of data collected at $\sqrt{s}$=13 TeV and $\sqrt{s}$=13.6 TeV with the ATLAS detector}", 
      author        = {ATLAS Collaboration},
      year          = {2025},
      eprint        = {2510.02857},
      archivePrefix = {arXiv},
      primaryClass  = {hep-ex},
      url           = {https://arxiv.org/abs/2510.02857},
      note          = {Submitted to Phys. Lett. B}
}

@misc{CMS_XYHZZbb,
      title         = "{Search for a new resonance decaying to a Higgs boson and a scalar boson in events with two b jets and two Z bosons in proton-proton collisions at $\sqrt{s}$ = 13.6 TeV}", 
      author        = {CMS Collaboration},
      year          = {2026},
      eprint        = {2602.18223},
      archivePrefix = {arXiv},
      primaryClass  = {hep-ex},
      url           = {https://arxiv.org/abs/2602.18223},
      note          = {Submitted to Phys. Rev. D}
}

@misc{HLLHC,
      title         = "{Highlights of the HL-LHC physics projections by ATLAS and CMS}", 
      author        = {ATLAS and CMS Collaborations},
      year          = {2025},
      eprint        = {2504.00672},
      archivePrefix = {arXiv},
      primaryClass  = {hep-ex},
      url           = {https://arxiv.org/abs/2504.00672}, 
}

\end{document}